\documentclass[conference]{IEEEtran}

\usepackage{graphicx}
\usepackage{color}
\usepackage{subfig}
\usepackage{cite,url}
\usepackage{comment}
\usepackage{enumerate}

\usepackage{amsmath}
\usepackage{amssymb}
\usepackage{amsfonts}
\usepackage{mathtools}
\usepackage{bm,bbm}
\usepackage{booktabs}

\usepackage{algorithm,algpseudocode}

\DeclareMathOperator*{\argmax}{arg\,max}

\renewcommand{\figurename}{Fig.}
\renewcommand{\tablename}{TABLE}
\newcommand{\sectionname}{Section}
\newcommand{\noindentbf}[1]{\noindent\textbf{#1}\quad}

\newcommand{\vspacelb}{\vspace{+0.3em}\\}

\renewcommand{\baselinestretch}{0.925}

\begin{document}

\title{
    ACK-Less Rate Adaptation for IEEE 802.11bc \\
    Enhanced Broadcast Services Using \\
    Sim-to-Real Deep Reinforcement Learning
}
\author{
    \IEEEauthorblockN{
        \normalsize Takamochi Kanda\IEEEauthorrefmark{1},
        \normalsize Yusuke Koda\IEEEauthorrefmark{1}\IEEEauthorrefmark{3},
        \normalsize Koji Yamamoto\IEEEauthorrefmark{1}\IEEEauthorrefmark{4}, and
        \normalsize Takayuki Nishio\IEEEauthorrefmark{1}\IEEEauthorrefmark{2}
    }
    \vspace{+0.5pt}
    \IEEEauthorblockA{
        \IEEEauthorrefmark{1}\small Graduate School of Informatics, Kyoto University,
        Yoshida-honmachi, Sakyo-ku, Kyoto, 606-8501, Japan \vspace{+0.5pt}\\
        \IEEEauthorrefmark{2}\small School of Engineering, Tokyo Institute of Technology,
        Ookayama, Meguro-ku, Tokyo, 158-0084, Japan \vspace{+0.5pt}\\
        \IEEEauthorrefmark{3}\small koda@imc.cce.i.kyoto-u.ac.jp,
        \IEEEauthorrefmark{4}\small kyamamot@i.kyoto-u.ac.jp
    }
}

\maketitle

\begin{abstract}

    In IEEE 802.11bc, the broadcast mode on wireless local area networks (WLANs),
    data rate control that is based on acknowledgement (ACK) mechanism
    similar to the one in the current IEEE 802.11 WLANs
    is not applicable because ACK mechanism is not implemented.
    This paper addresses this challenge by proposing
    ACK-less data rate adaptation methods
    by capturing non-broadcast uplink frames of STAs.
    In IEEE 802.11bc, an use case is assumed,
    where a part of STAs in the broadcast recipients
    is also associated with non-broadcast APs,
    and such STAs periodically transmit uplink frames including ACK frames.
    The proposed method is based on the idea that by overhearing such uplink frames,
    the broadcast AP surveys channel conditions at partial STAs,
    thereby setting appropriate data rates for the STAs.
    Furthermore, in order to avoid reception failures in a large portion of STAs,
    this paper proposes deep reinforcement learning (DRL)-based
    data rate adaptation framework that uses a sim-to-real approach.
    Therein, information of reception success/failure at broadcast recipient STAs,
    that could not be notified to the broadcast AP in real deployments,
    is made available by simulations beforehand,
    thereby forming data rate adaptation strategies.
    Numerical results show that utilizing overheard uplink frames of recipients
    makes it feasible to manage data rates in ACK-less broadcast WLANs,
    and using the sim-to-real DRL framework
    can decrease reception failures.

\end{abstract}
\IEEEpeerreviewmaketitle

\section{Introduction}
\label{sec:introduction}

The IEEE 802.11 working group is currently working on
the standardization of IEEE 802.11bc~\cite{11bcFunctionalRequirementsDocument}.
IEEE 802.11bc provides broadcasting on wireless local area networks (WLANs)
termed enhanced broadcast services (eBCS)~\cite{11bcPAR, 11bcEBCSDemonstration},
which is applied to services of common contents for a large number of people
at specific locations, such as eSports virtual reality videos,
lecture room slide distributions,
and emergency broadcasting~\cite{11bcUseCaseDocument}.

In IEEE 802.11bc, acknowledgement (ACK) mechanisms
cannot be implemented between the eBCS access point (AP)
and eBCS recipient stations (STAs)~\cite{11bcBroadcastingOnWLAN}.
Hence, the eBCS APs are not notified whether the STAs have received
the broadcast frames successfully or not.
This renders it quite challenging for the eBCS APs to control
their parameters, e.g., data rate and transmit power,
adaptively to channel conditions or distances to eBCS recipient clusters.
Meanwhile, in view of a deployment scenario of eBCS,
this challenge should be addressed for the following two reasons.
First, being different from other wireless broadcast systems, e.g., TV broadcasting,
WLANs are generally self-deployed.
This indicates that the eBCS APs are deployed in various locations without
as much concise installation designs as current wireless broadcast systems.
Hence, for ease of deployment, as in the current WLANs,
autonomous adaptations of parameters of APs
for various installation locations are required.
Secondly, due to being different from TV broadcasting,
channel conditions for the recipient STAs vary for each events, e.g.,
for eSports event venues and lecture room.
This mandates eBCS APs to autonomously adapt their parameters
to various channel conditions to successfully deliver the eBCS contents.

We address this challenge by developing
an ACK-less rate adaptation method
harnessing the uplink frames from the STAs associated with non-eBCS APs,
which is surrounded by other eBCS recipients.
In view of eBCS application,
we can easily envisage situations where
some of the STAs are associated with other non-eBCS APs,
and they transmit uplink frames to the non-eBCS AP
as in the current application of WLANs.
Hence, by overhearing the uplink frames from such STAs,
eBCS APs capture the channel conditions or distances from the clusters of
the surrounding eBCS recipients, thereby setting the data rate adaptively.
We refer to this idea as
\textit{frame overhearing-based recipient estimation} (FO-RE).
To validate the feasibility of the idea, as an example,
we will consider that an eBCS AP measures received signal strength (RSS)
of the overheard uplink frames from the STAs associated with the non-eBCS APs,
thereby measuring the distance for the STAs.
Based on this, the eBCS AP selects appropriate data rates to
not only the STAs that are transmitting uplink frames,
but also, the surrounding eBCS recipients.

Nevertheless, since not all eBCS recipients are transmitting uplink frames,
the eBCS AP cannot manage data rates for the STAs that are not transmitting
uplink frames.
More concretely, the RSS of the overheard uplink frames from the STAs
do not represent the distances to all eBCS recipients.
This leads to frequent reception failures at residual eBCS recipients
that are not transmitting uplink frames
whose distances from the eBCS AP are inaccurately estimated.

To solve this issue, we provide a sim-to-real
deep reinforcement learning (DRL)~\cite{mnih2013playing} framework
for ACK-less data rate adaptation in eBCS systems.
This framework consists of two phases, i.e.,
\textit{learning phase} and \textit{application phase}.
In the learning phase, the eBCS AP learns a data rate control policy using DRL,
where the aforementioned RSS information is used as a state.
Therein, a key idea is that the information of the reception success/failure
at all eBCS recipients,
which could not be obtained in a real deployment,
is made available by simulations.
Based on this information,
the reward is designed so that the reception failures are alleviated
in the subsequent real applications.
In the application phase, the learned policy is applied to the real environment,
where reception success/failure information is not available.
By using DRL, the eBCS AP learns appropriate data rates
in view of the distance from the STA while avoiding reception failures.

The contributions of this paper are as follows:
\begin{itemize}
    \item To the best of our knowledge, this is the first work to develop
          an \textit{ACK-less rate adaptation method for broadcast WLANs}.
          The key idea to overcome the challenge of adapting the data rate
          in the ACK-less situation is to overhear uplink feedbacks
          in non-broadcast WLANs overlaid by broadcast WLANs,
          which we term as FO-RE.
          More specifically, we leverage the overheard uplink frames
          from the STAs associated with the non-broadcast APs, and
          measures the channel conditions for the eBCS recipients;
          thereby selecting the data rates adaptively.
    \item We further discuss a problem-specific challenge
          of our proposed ACK-less rate adaptation method.
          To solve this, we propose a sim-to-real learning-based framework.
          More specifically, the captured channel conditions at
          do not necessarily represent the distances for all eBCS recipient STAs,
          causing frequent reception failures at partial STAs.
          The idea to alleviate these failures is to leverage
          hypothetical feedbacks on the reception success/failure at all recipient STAs of eBCS
          yielded in a simulation, based on which we design an objective function, i.e., reward,
          to represent the effectiveness of the data rate selection
          for the clusters of recipient STAs of eBCS.
          By learning a data rate selection policy to maximize the reward via DRL,
          the eBCS AP can select appropriate data rates for the recipients
          while alleviating reception failures.
\end{itemize}

The remainder of this paper is organized as follows:
In \sectionname~\ref{sec:system_model}, we provide a system model.
Secondly, in \sectionname~\ref{sec:proposed_method_rule_based},
we propose an ACK-less data rate adaptation method harnessing
the uplink frames from the STAs associated with non-eBCS APs,
and in \sectionname~\ref{sec:proposed_method_learning_based},
we propose a sim-to-real learning-based method using DRL.
Subsequently in \sectionname~\ref{sec:performance_evaluation},
we describe a performance evaluation of the aforementioned proposed methods.
Finally, we provide our concluding remarks in \sectionname~\ref{sec:conclusion}.

\section{System Model}
\label{sec:system_model}

\begin{figure}[t]
    \centering
    \includegraphics[width=0.9\columnwidth]{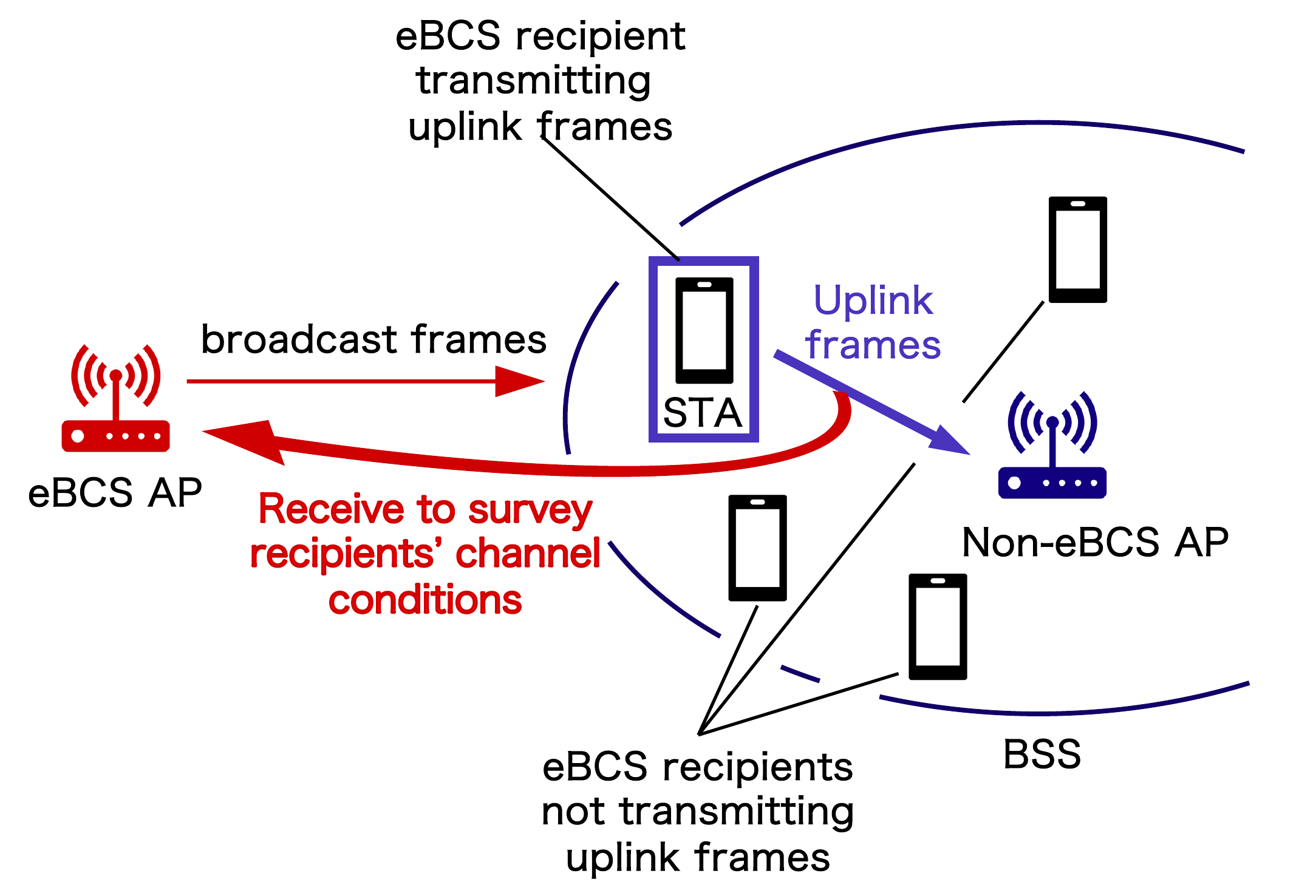}
    \caption{
        An example of the considered broadcast communication system.
        Uplink frames to non-eBCS APs are also received by
        an eBCS AP to survey channel conditions between the eBCS AP and STAs.
    }
    \label{fig:system_model}
\end{figure}

We consider a broadcast communication system, where one eBCS AP,
$I$~non-eBCS APs, and multiple eBCS recipient STAs are deployed in
a two-dimensional area~$\mathcal{W}\subset\mathbb{R}^{2}$.
Note that the proposed framework can be easily
extended to a scenario with multiple eBCS APs.
The eBCS AP periodically transmits broadcast frames to STA clusters.
The STA clusters are composed of $I$~basic service sets (BSSs),
and BSS~$i\in\{1, 2, \dots, I\}$ consists of one non-eBCS AP and $N_{i}$~recipient STAs,
where the total number of recipient STAs is denoted by $N = \sum_{i=1}^{I}N_{i}$.
We assume that these APs and STAs are densely deployed.
Hence, we focus on the BSSs far from the eBCS AP,
where partial STAs can fail to receive broadcast frames.
An example of the system model is depicted in \figurename~\ref{fig:system_model}.
We consider that the eBCS AP can receive uplink frames
from the STAs to survey the channel conditions
between the eBCS AP and the recipient STAs.

\section{ACK-Less Rate Adaptation Harnessing Uplink Frames from Non-Broadcast WLANs}
\label{sec:proposed_method_rule_based}

The objective of this section is to describe the procedure
of managing data rates in ACK-less broadcast.
The main idea is to utilize the overheard uplink frames from STAs,
which are surrounded by eBCS recipients,
associated with the non-eBCS APs.
We refer to this idea as FO-RE.
For demonstration, a rule-based method \textsf{FO-RE-Rule} is proposed in this section,
where the eBCS AP leverages the RSS of the frames.
In the proposed \textsf{FO-RE-Rule},
the eBCS AP implements the following two procedures in one step:
\vspacelb
\noindentbf{RSS Observation.}
Firstly, the eBCS AP receives the uplink frames of $m$~STAs for each step,
i.e., a step is defined by the time of receiving $m$~frames of the eBCS AP.
Note that the eBCS AP may not be always able to receive all the STA frames,
i.e., sometimes $m<N$, and the $m$~STAs are different for each step.
The eBCS AP observes the RSSs of the uplink frames
from the $m$~STAs,
and the space of the RSS vector of the received frames is expressed as:
\begin{equation}
    \mathcal{P} \coloneqq
    \underbrace{\mathcal{J}\times\mathcal{J}\times\dots\times\mathcal{J}}_{m},
\end{equation}
where $\mathcal{J}\subset\mathbb{R}$ is the space of RSS in $\mathrm{dBm}$.
Denoting each RSS by $p_{t}^{(j)}\in\mathcal{J}$ for $j=1, 2, \dots , m$,
an RSS vector~$\bm{p}_{t}$ sampled from $\mathcal{P}$ in step~$t$ is denoted by:
\begin{equation}
    \bm{p}_{t} \coloneqq
    \left(
    p_{t}^{(1)}, p_{t}^{(2)}, \dots , p_{t}^{(m)}
    \right).
\end{equation}
\vspacelb
\noindentbf{Data Rate Selection.}
Secondly, based on the observed RSSs~$\bm{p}_{t}$, the eBCS AP determines
a data rate~$a_{t}$ for each step~$t$
so that, at least, the above $m$~STAs can receive broadcast frames.
When determining the data rate,
the eBCS AP first calculates the path loss between the AP and each STA as:
\begin{equation}
    L_{t}^{(j)} \coloneqq P_{\mathrm{STA}} - p_{t}^{(j)}.
\end{equation}
Based on this, the eBCS AP estimates a received SNR of broadcast frames at each STA,
which is calculated using the equation below:
\begin{equation}
    \gamma_{t}^{(j)} \coloneqq P_{\mathrm{eBCS}} - L_{t}^{(j)} - P_{\mathrm{n}},
\end{equation}
where $P_{\mathrm{n}}$ is a noise power in $\mathrm{dBm}$
and $P_{\mathrm{STA}}, P_{\mathrm{eBCS}}$ are
the transmit powers of the STAs and the eBCS AP respectively.
Subsequently, the data rate in step~$t$ is determined as follows:
\begin{equation}
    a_{t} = \max_{a}\{\, a\mid\min_{j}\gamma_{t}^{(j)}\geq\gamma_{\mathrm{req}}(a) \,\},
\end{equation}
where $\gamma_{\mathrm{req}}(a)$ is the SNR required
to receive the broadcast frames transmitted in the data rate~$a$ successfully.
Note that the eBCS AP is assumed to know $P_{\mathrm{STA}}$
by referring to standard settings of the transmit power for IEEE 802.11 WLANs.

\section{Sim-to-Real Deep Reinforcement Learning Framework for ACK-Less Rate Adaptation}
\label{sec:proposed_method_learning_based}

In this section, we propose a sim-to-real learning-based
data rate management framework.
The main objective of this section is to avoid the
reception failures at not only the above $m$~STAs
that are transmitting the uplink frames,
but also at the residual $N-m$~STAs that are not transmitting those frames.
First, we describe two phases for sim-to-real learning,
i.e., the learning phase and application phase.
Secondly, to avoid reception failures at the residual $N-m$~STAs,
we propose a sim-to-real DRL framework for ACK-less
rate adaptation for broadcasting.
This framework also uses the idea of FO-RE.
We refer to the proposed method in this section as \textsf{FO-RE-DRL}.

\subsection{Learning Phase and Application Phase}

The proposed \textsf{FO-RE-DRL} consists of two phases,
i.e., the learning phase and application phase.
In the learning phase, using DRL, the eBCS AP obtains
a policy for ACK-less data rate adaptation,
where the eBCS AP implements steps, defined as
the consequence of the RSS and BSS identifier (BSSID)
observation and data rate selection
as well as in \textsf{FO-RE-Rule}.
Importantly, in this phase, the eBCS AP can observe
the information of the reception success/failure
not only at the $m$~STAs that are transmitting the uplink frames,
but also, at the residual $N-m$~STAs that are not transmitting those frames.
The availability of this information is the main reason
for the proposed \textsf{FO-RE-DRL} alleviating
the reception failure at the $N-m$~residual STAs.

In the application phase, the eBCS AP applies the policy obtained
in the learning phase to the real environment.
Note that the eBCS AP implements steps similarly to the learning phase,
but cannot observe the reception success/failure information in this phase.
This is because in view of the eBCS applications,
it is difficult for the eBCS AP to obtain the success/failure
information in the real environment.

\subsection{Sim-to-Real Deep Reinforcement Learning Framework}

\begin{figure}[t]
    \centering
    \includegraphics[width=\columnwidth]{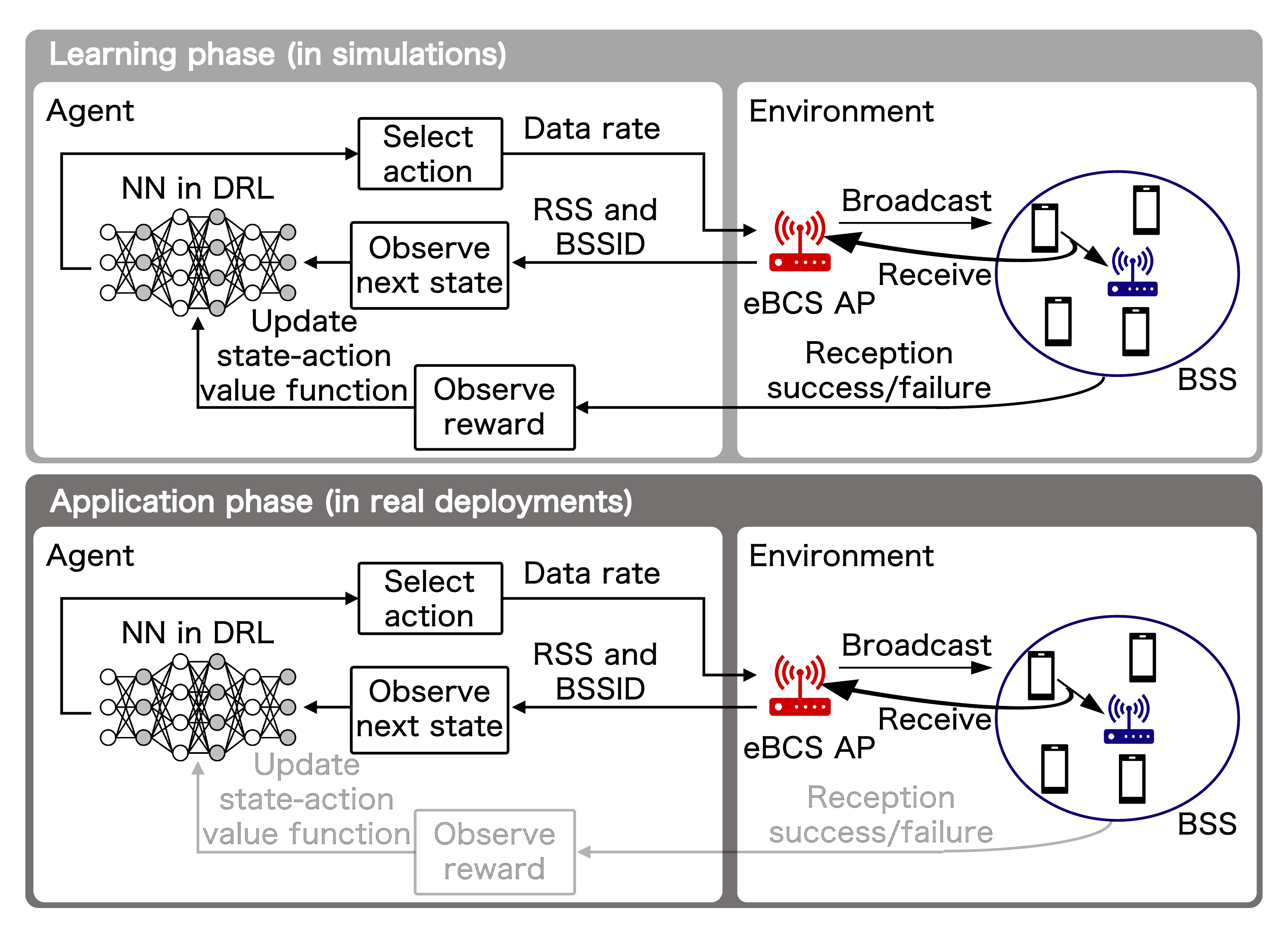}
    \caption{
        Learning phase and application phase flows.
        The learning phase is implemented in simulations,
        where reception success/failure information
        is notified to the eBCS AP,
        which is not notified in the application phase
        implemented in real deployments.
    }
    \label{fig:learning_phase_and_application_phase_flow}
\end{figure}

The proposed \textsf{FO-RE-DRL} uses DRL algorithm.
The agent is defined as the data rate controller of the eBCS AP
and selects the data rates following the policy~$\pi$.
Assuming that an interaction between the agent and the environment
is described by the Markov decision process (MDP),
the agent observes a state and a reward and selects an action
in each step~\cite{RL}.
We define the state, action, and reward as follows:
\vspacelb
\noindentbf{State.}
We use the RSSs and BSSIDs of the uplink frames from the $m$~STAs
associated with the non-eBCS APs,
where the $m$~STAs are different for each step.
In \textsf{FO-RE-DRL},
the eBCS AP observes the BSSIDs of overheard uplink frames
to distinguish which BSS each STA belongs to.
Given this, state space~$\mathcal{S}$ is expressed as follows:
\begin{equation}
    \mathcal{S} \coloneqq \mathcal{P}\times\mathcal{Q}, \quad
    \mathcal{Q} \coloneqq
    \underbrace{\mathcal{I}\times\mathcal{I}\times\dots\times\mathcal{I}}_{m},
\end{equation}
where $\mathcal{I}$ is the space of BSSIDs.
Denoting the BSSIDs by $q_{t}^{(j)}\in\mathcal{I}$ for $j = 1, 2, \dots , m$,
the state in step~$t$ is expressed by:
\begin{equation}
    \bm{s}_{t} = \left(
    p_{t}^{(1)}, p_{t}^{(2)}, \dots , p_{t}^{(m)},
    q_{t}^{(1)}, q_{t}^{(2)}, \dots , q_{t}^{(m)}
    \right).
\end{equation}
\vspacelb
\noindentbf{Action.}
The action is defined as selecting a data rate.
The action space~$\mathcal{A}$ is defined as follows:
\begin{equation}
    \mathcal{A} \coloneqq \{A_{1}, A_{2}, \dots , A_{K}\},
\end{equation}
where $A_{1}, A_{2}, \dots , A_{K}$ represent the available rates in $\mathrm{Mbit/s}$.
The action sampled from $\mathcal{A}$ in step~$t$ is denoted by $a_{t}$.
\vspacelb
\noindentbf{Reward.}
Aiming to avoid reception failures at the eBCS recipients,
we define the reward $r_{t+1}$ as
\begin{equation}
    r_{t+1} \coloneqq
    \begin{cases}
        - (a_{t}/a_{\mathrm{max}})(1 - n_{t}/N),
         & n_{t} < N; \\
        a_{t}/a_{\mathrm{max}},
         & n_{t} = N,
    \end{cases}
\end{equation}
where $n_{t}$ is the number of recipient STAs
that succeed in receiving the broadcast frames
transmitted by the eBCS AP in step~$t$,
and $a_{\mathrm{max}}$ is the available maximum data rate.
We separate the reward by $n_{t} = N$, so that the reward
is always less than zero when $n_{t} < N$
whereas it is more than zero when $n_{t} = N$.
Through this design, the agent will be sensitive to the reception failures
due to its negative reward;
and therefore, it will avoid these reception failures.
Note that the reward is in $[-1, 1]$, divided by $a_{\mathrm{max}}$,
which stabilizes the learning.
\vspacelb
In the learning phase, the agent can observe the reward,
meanwhile, in the application phase, cannot observe it.
The flows of the phases are shown in
\figurename~\ref{fig:learning_phase_and_application_phase_flow}.

\section{Performance Evaluation}
\label{sec:performance_evaluation}

In this section, we evaluate the performances of the proposed
\textsf{FO-RE-Rule} and \textsf{FO-RE-DRL},
by conducting test simulations imitating the real applications,
since we aim to evaluate the effectiveness of FO-RE and
the proposed sim-to-real learning-based framework.
We firstly describe the detailed settings for the simulation environment
and DRL agent in \textsf{FO-RE-DRL}.
Secondly, we explain the compared methods in this evaluation.
Subsequently, we provide a simulation procedure composed of
the learning phase simulation and test simulation
imitating the application phase.
Finally, we demonstrate the numerical results and discuss them.

\subsection{Environment and DRL Agent Settings}

\noindentbf{Environment Settings.}
We detail the setting for the simulation environment.
As an example, we use some of the data rates
defined in IEEE 802.11ax~\cite{11axFunctionalRequirementsDocument},
where the action space~$\mathcal{A}$ is expressed as
\begin{equation}
    \mathcal{A} = \{8.6, 51.6, 103.2, 143.4\}.
\end{equation}

Due to carrier sensing, it is assumed that
no frame collisions occur among the eBCS AP, the non-eBCS AP,
and the recipient STAs.
In these, reception failures are only caused by
signal-to-noise power ratio (SNR)
that is lower than those required for the recipients to receive
the broadcast frames, and this depends on the selected data rates of the eBCS AP.
The required SNR~$\gamma_{\mathrm{req}}$ in data rate $a$ is defined as follows:
\begin{equation}
    \gamma_{\mathrm{req}}(a) \coloneqq 2^{a/W} - 1,
    \label{eq:req_snr}
\end{equation}
where $W$ is the bandwidth of the system.
This requirement is derived from Shannon's noisy-channel coding theorem
stating that the upper limit of the data rate for error-free
communication is the channel capacity.
Other detail settings are shown in \tablename~\ref{tbl:environment_settings}.
Note that the learning phase and application simulations in \textsf{FO-RE-DRL} use
the same environment settings, excluding whether the information of
reception success/failure is notified to the eBCS AP or not.
\vspacelb
\noindentbf{DRL Agent Settings.}
In \textsf{FO-RE-DRL},
following the DRL algorithm,
deep Q-network (DQN)~\cite{mnih2015human},
the agent trains a neural network (NN) and learns the optimal
state-action value function~$Q^{\star}(s, a)$,
which is defined by the expectation of the cumulative sum of the discounted reward.
The optimal policy is obtained by $\pi^{\star} \coloneqq \argmax_{a} Q^{\star}(s, a)$.
The agent selects the action following $\varepsilon$-greedy policy in the learning phase,
and following greedy policy in the application phase.
The detail settings of the DQN agent in the learning phase are shown
in \tablename~\ref{tbl:dqn_agent_settings_in_learning_phase}.
The NN in DQN is composed of six fully connected layers,
where each hidden layer consists of 64 units,
and uses rectified linear unit (ReLU) activation.

\begin{table}[t]
    \centering
    \caption{Environment Settings}
    \begin{tabular}[t]{cc}
        \toprule
        Region~$\mathcal{W}$                   & $300\,\mathrm{m}\times300\,\mathrm{m}$          \\
        Carrier frequency~$f_{\mathrm{c}}$     & $5\,\mathrm{GHz}$                               \\
        Bandwidth~$W$                          & $20\,\mathrm{MHz}$                              \\
        Break point distance~$d_{\mathrm{BP}}$ & $10\,\mathrm{m}$                                \\
        Transmit power of the eBCS AP~$P_{\mathrm{eBCS}}$
                                               & $10\,\mathrm{mW}$                               \\
        Transmit power of the STAs~$P_{\mathrm{STA}}$
                                               & $10\,\mathrm{mW}$                               \\
        Number of non-eBCS APs~$I$             & $2$                                             \\
        Total number of STAs~$N$               & $100$                                           \\
        \begin{tabular}{c}
            Number of received STAs' uplink \vspace{-2pt} \\
            frames in one step~$m$
        \end{tabular}
                                               & $5$                                             \\
        Path loss model                        & Indoor model in \cite{11axChannelModelDocument} \\
        \bottomrule
    \end{tabular}
    \label{tbl:environment_settings}
\end{table}

\begin{table}[t]
    \centering
    \caption{DQN Agent Settings in Learning Phase}
    \begin{tabular}[t]{cc}
        \toprule
        Number of episodes             & $10{,}000$                               \\
        Number of steps in one episode & $100$                                    \\
        Policy                         & $\varepsilon$-greedy ($\varepsilon=0.3$) \\
        Learning rate                  & $0.0001$                                 \\
        Discount factor                & $0$                                      \\
        Batch size                     & $32$                                     \\
        Loss function                  & Huber loss~\cite{huber1992robust}        \\
        Optimizer                      & Adam~\cite{kingma2014adam}               \\
        Replay buffer capacity         & $10{,}000$                               \\
        \bottomrule
    \end{tabular}
    \label{tbl:dqn_agent_settings_in_learning_phase}
\end{table}

\subsection{Compared Methods}

In order to confirm that utilizing the overheard uplink frames
can make ACK-less rate management feasible,
we compare the proposed methods with the baseline method \textsf{MinRate}.
In \textsf{MinRate}, the eBCS AP always selects the minimum rate~$8.6\,\mathrm{Mbit/s}$,
and hence, this method does not need FO-RE.
Comparison of all methods is summarized
in \tablename~\ref{tbl:comparison_of_methods}.

\begin{table}[t]
    \centering
    \caption{Comparison of All Methods}
    \begin{tabular}[t]{ccc}
        \toprule
        Methods
         & w/ FO-RE?
         & \begin{tabular}{c}
            w/ sim-to-real \\ learning?
        \end{tabular}                \\
        \midrule
        \textsf{MinRate}
         & No                         & No           \\
        \textbf{proposed~1:} \textsf{FO-RE-Rule}
         & \textbf{Yes}               & No           \\
        \textbf{proposed~2:} \textsf{FO-RE-DRL}
         & \textbf{Yes}               & \textbf{Yes} \\
        \bottomrule
    \end{tabular}
    \label{tbl:comparison_of_methods}
\end{table}

\subsection{Simulation Procedure}

In order to evaluate the performances of each method,
we implement the test simulations imitating the real environment.
In the test simulations, the episodes, which are
defined as a particular sequence of $100$ steps, are implemented.
The total number of episodes is set to $1{,}000$.
For each episode, the eBCS AP, the non-eBCS APs,
and the STAs are randomly deployed.
As an example, we deploy the STAs according to the daughter processes of
Thomas process~\cite{thomas1949generalization},
regarding the non-eBCS APs as parent points.
Note that the learning phase simulation in \textsf{FO-RE-DRL}
is conducted beforehand,
where the deployments of the APs and STAs are
the same as those in the test simulations.

Our goal is to manage the data rates in the ACK-less environment, and
to make all the STAs succeed in receiving the broadcasted frames from the eBCS AP
while enhancing the throughput.
Hence, we evaluate the performances of each method
with two metrics, which are aggregated throughput and success ratio calculated
on all the eBCS recipients.

\begin{figure}[t]
    \centering
    \includegraphics[width=0.9\columnwidth]{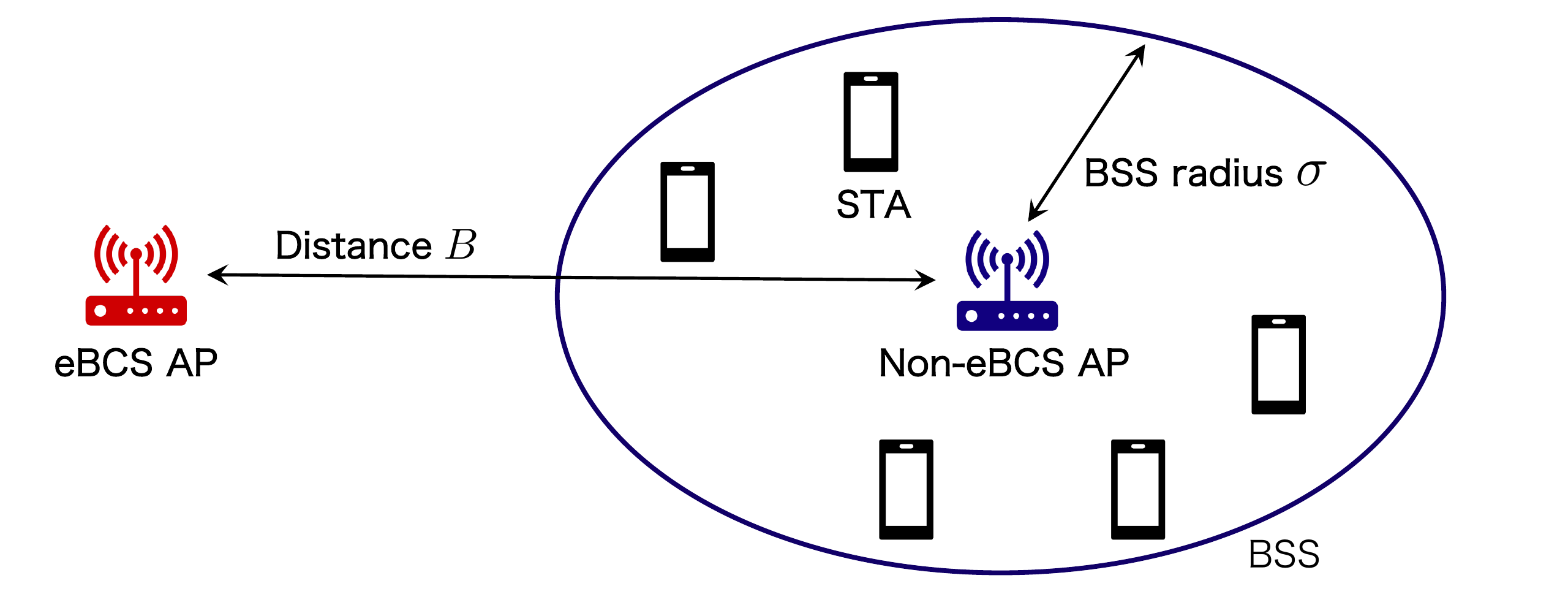}
    \caption{Distance~$B$ and BSS radius~$\sigma$.}
    \label{fig:dmax_and_scale_imiplication}
\end{figure}

In the test simulations,
to evaluate the performances against various deployments of STAs,
we vary the following two parameters:
distance~$B$ and BSS radius~$\sigma$.
The distance~$B$ is defined as the distance between
the eBCS AP and the farthest non-eBCS AP, which is calculated by:
\begin{equation}
    B \coloneqq
    \max_{i}|\bm{x}_{\mathrm{eBCS}} - \bm{x}_{i}|,
\end{equation}
where the position of the eBCS AP and the non-eBCS AP~$i$ is denoted by
$\bm{x}_{\mathrm{eBCS}}$ and $\bm{x}_{i}$ respectively.
The BSS radius~$\sigma$ is defined as the standard deviation of
the positions of the STAs.
\figurename~\ref{fig:dmax_and_scale_imiplication} shows
the implications of the distance~$B$ and the BSS radius~$\sigma$.

\subsection{Results}

\begin{figure}[t]
    \centering
    \subfloat[Aggregated throughput.]{\includegraphics[width=0.465\columnwidth]{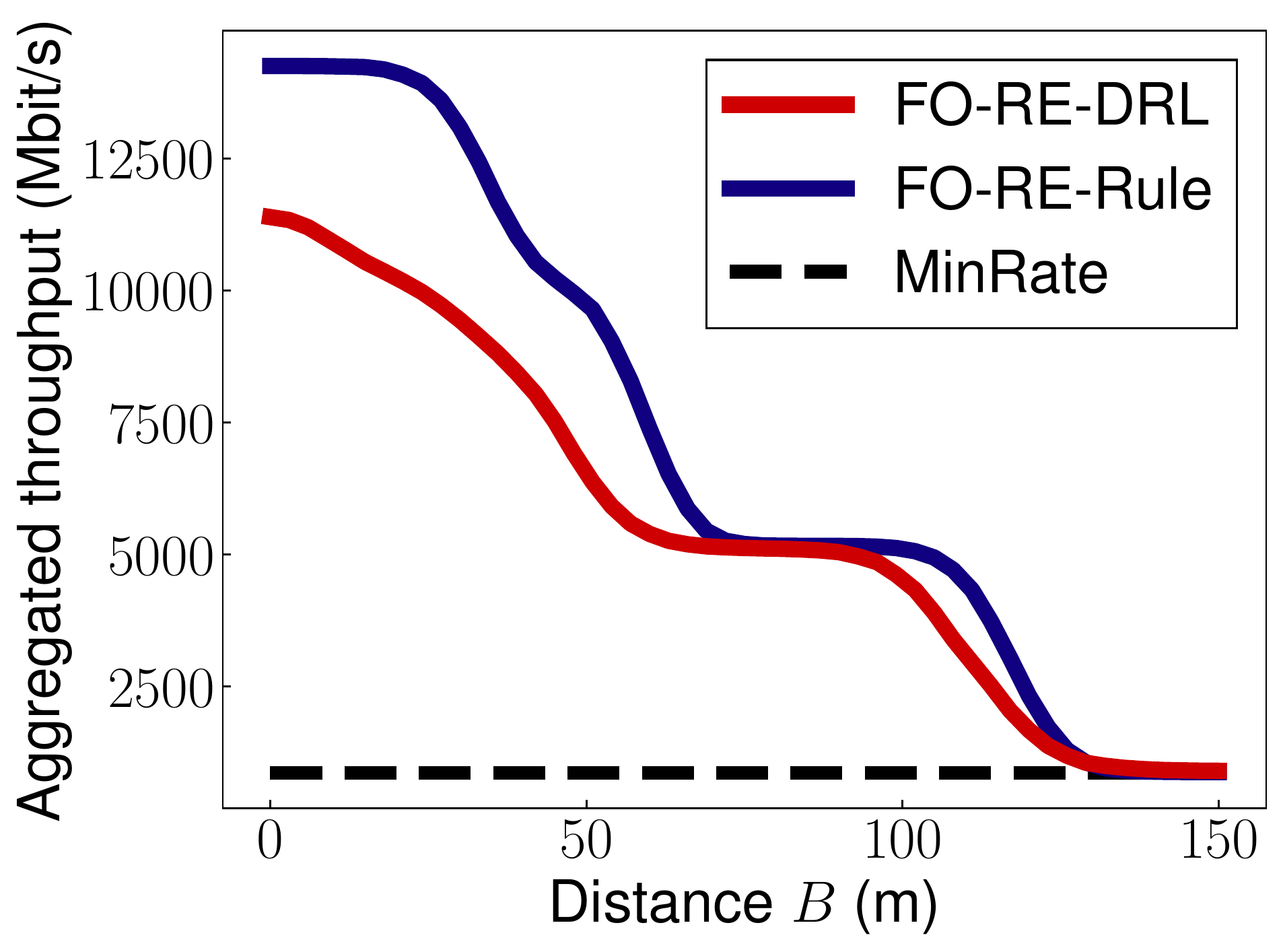}}
    \subfloat[Success ratio.]{\includegraphics[width=0.465\columnwidth]{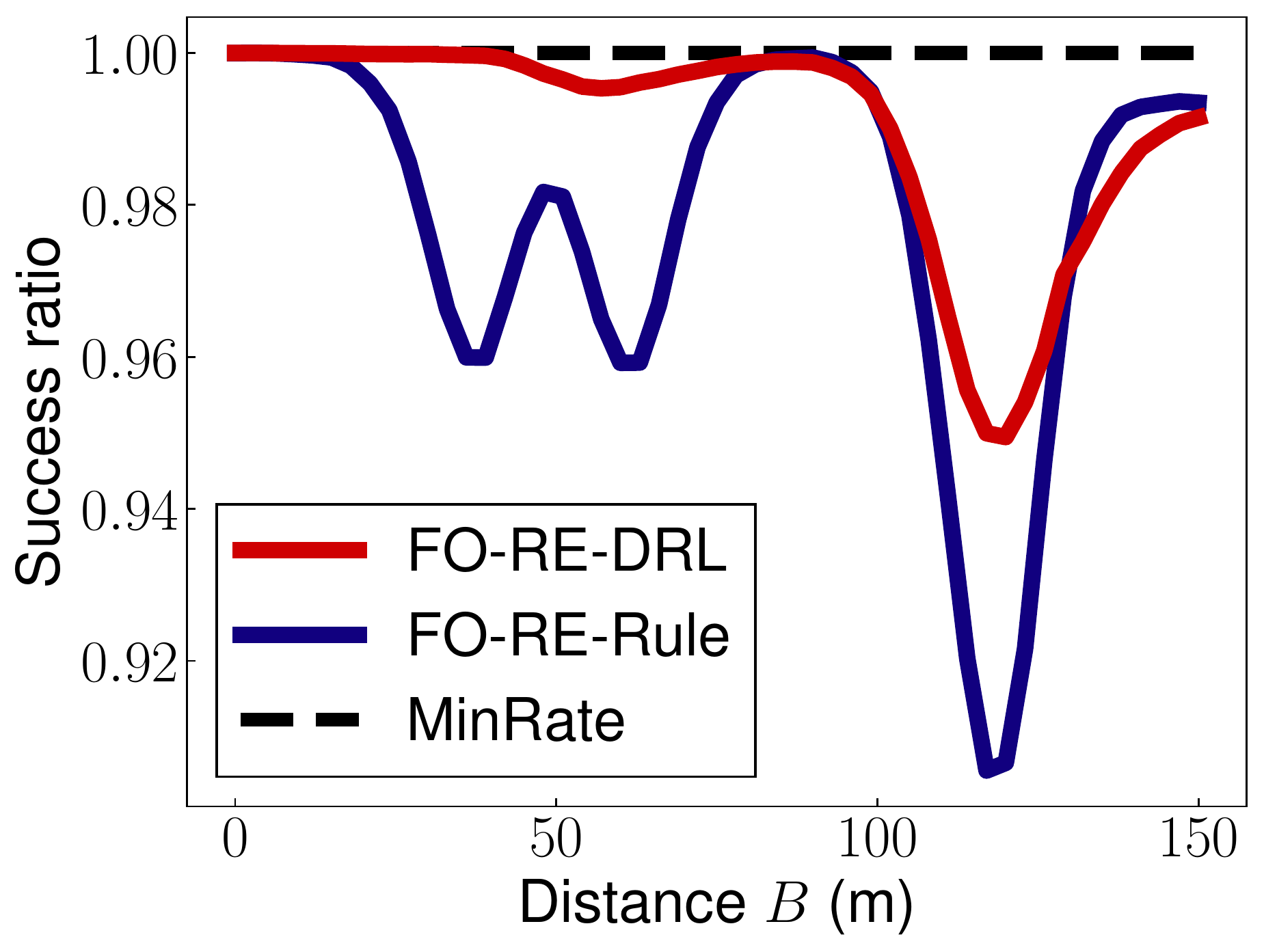}}
    \caption{
        Aggregated throughput and success ratio plotted against distance~$B$.
        BSS radius~$\sigma$ is fixed to $10\,\mathrm{m}$.
    }
    \label{fig:dmax_mean_throughput_dmax_mean_rsuccess}
\end{figure}

\begin{figure}[t]
    \centering
    \subfloat[Aggregated throughput.]{\includegraphics[width=0.465\columnwidth]{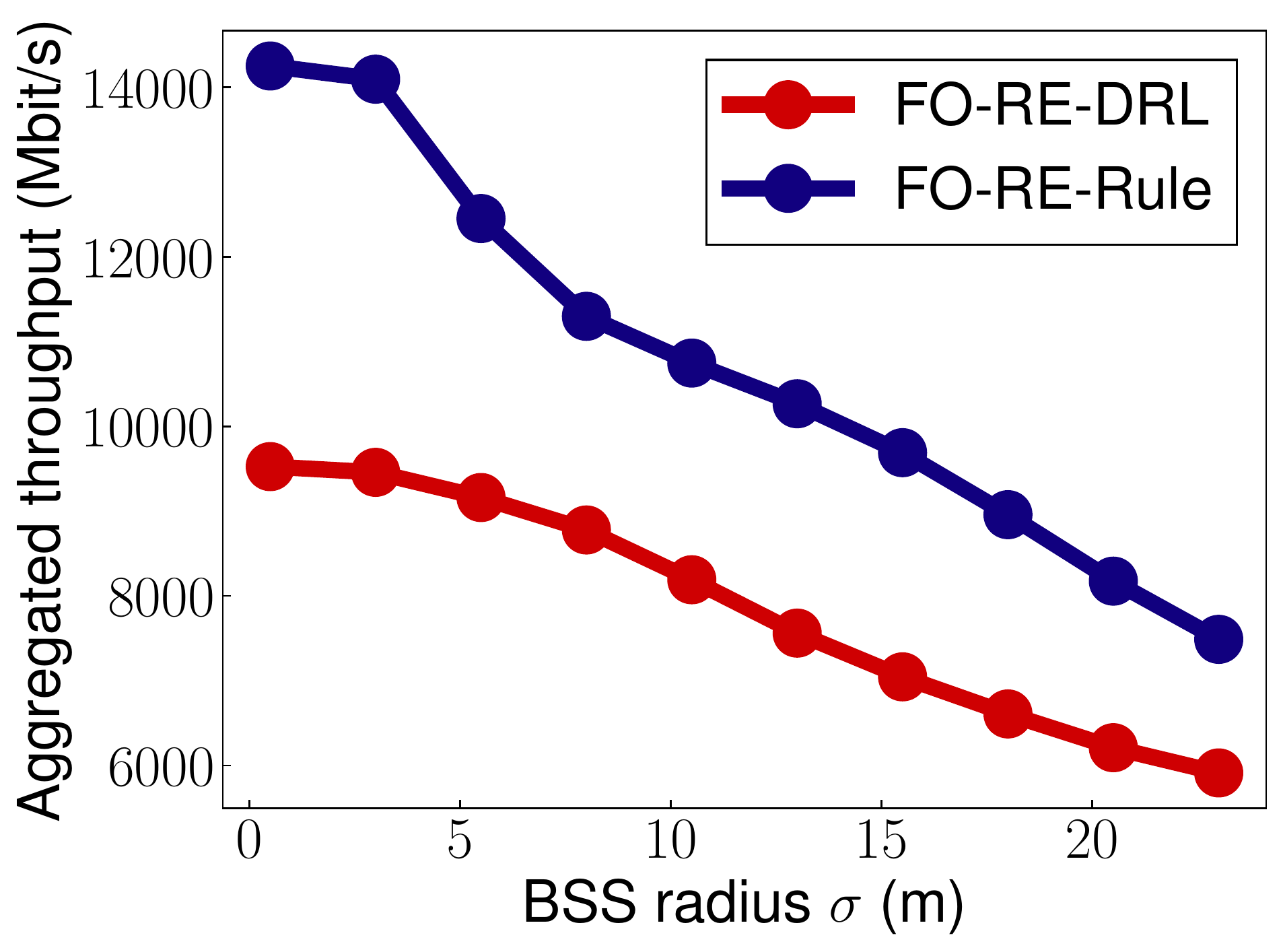}}
    \subfloat[Success ratio.]{\includegraphics[width=0.465\columnwidth]{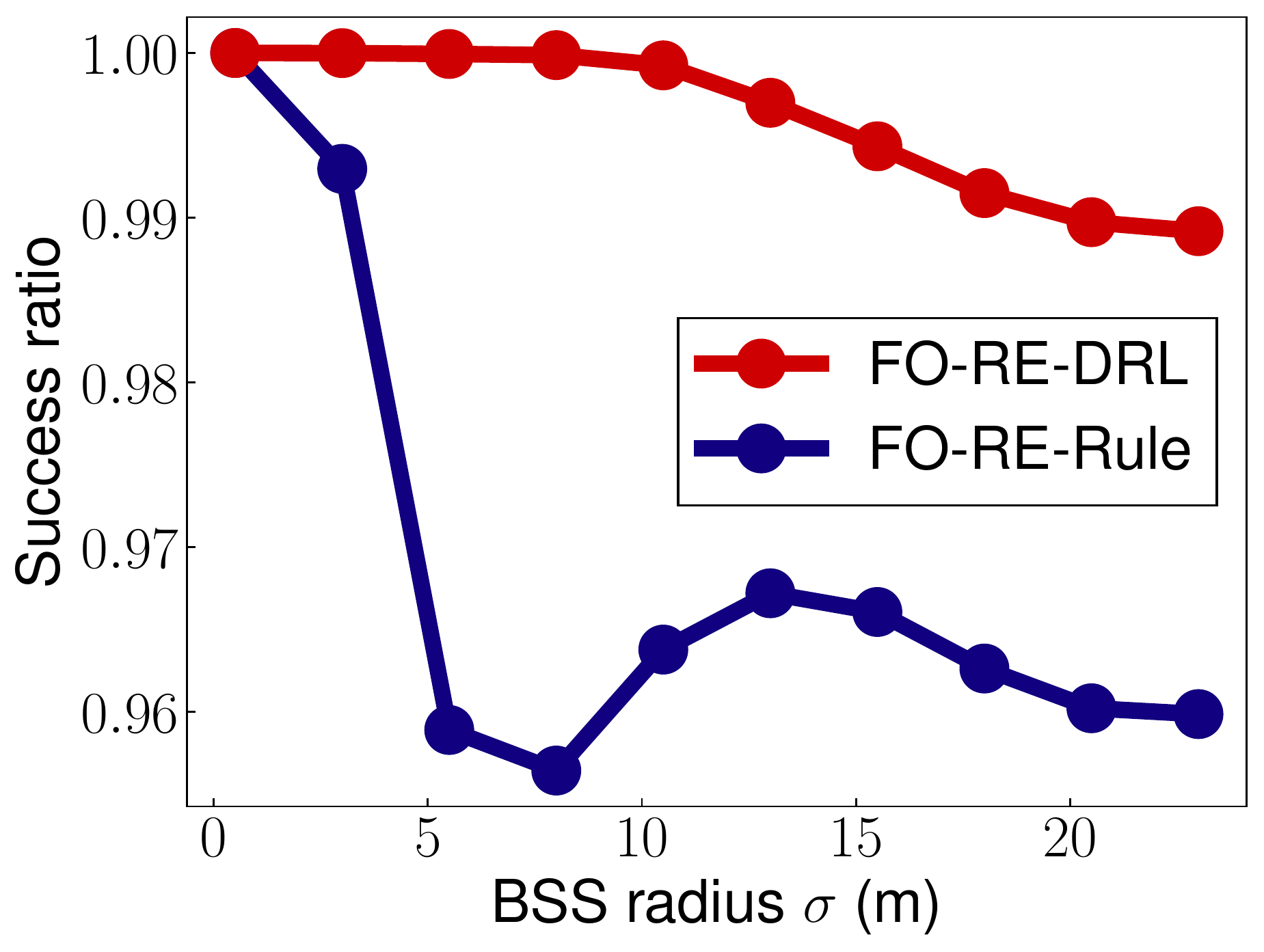}}
    \caption{
        Aggregated throughput and success ratio plotted against BSS radius~$\sigma$.
        Distance~$B$ is fixed to $40\,\mathrm{m}$.
    }
    \label{fig:scale_mean_throughput_scale_mean_rsuccess}
\end{figure}

\noindentbf{Evaluation Against Distance.}
\figurename~\ref{fig:dmax_mean_throughput_dmax_mean_rsuccess}
shows the advantage of utilizing the information of
the overheard uplink frames from the STAs associated with the non-eBCS APs.
The proposed \textsf{FO-RE-Rule} and \textsf{FO-RE-DRL}
achieve higher throughput than those in compared methods,
while maintaining high success ratio as in \textsf{MinRate}.
In \figurename~\ref{fig:dmax_mean_throughput_dmax_mean_rsuccess}(a),
when the distance~$B$ is short, the eBCS AP can
transmit at high data rates in the proposed
\textsf{FO-RE-Rule} and \textsf{FO-RE-DRL},
where these methods achieve high throughput.
\figurename~\ref{fig:dmax_mean_throughput_dmax_mean_rsuccess}(b)
reveals that the eBCS AP transmits in the data rates
which can cover almost all the STAs
in \textsf{FO-RE-Rule} and \textsf{FO-RE-DRL},
where the success ratio is kept high.
This implies that the idea of FO-RE makes ACK-less rate adaptation feasible.
\vspacelb
\noindentbf{Evaluation Against BSS Radius.}
The result in \figurename~\ref{fig:scale_mean_throughput_scale_mean_rsuccess}
confirms the feasibility of avoiding reception failures
in ACK-less broadcast by using the proposed sim-to-real DRL framework.
\textsf{FO-RE-DRL} demonstrates higher success ratio than \textsf{FO-RE-Rule}
even when the BSS radius is long.
This is because the eBCS AP transmits at slightly lower rates
in the proposed \textsf{FO-RE-DRL} in order to avoid causing reception failures.
In other words, \textsf{FO-RE-DRL} prioritizes achieving
a higher success ratio rather than enhancing throughput.
This difference is caused by the size of the BSSs.
In \textsf{FO-RE-Rule}, the eBCS AP selects the data rate
that can only cover the STAs that are transmitting uplink frames.
Meanwhile, the proposed \textsf{FO-RE-DRL} learns the expected value of the reward,
which also reflects reception success/failure at the STAs
that are not transmitting uplink frames
whose distances from the eBCS AP are inaccurately estimated,
and therefore, selects the data rates to cover all the recipient STAs.
This implies that reception failures are alleviated
by using the proposed sim-to-real learning-based framework.

\section{Conclusion}
\label{sec:conclusion}

In this paper, we developed ACK-less rate adaptation methods
for an eBCS AP.
These methods are based on the idea of utilizing the overheard uplink frames
from the STAs associated with the non-eBCS APs.
As an example, we proposed a rule-based method using the RSS of the overheard frames.
The numerical results demonstrated that utilizing the overheard frames makes it
feasible for the eBCS AP to manage the data rates in ACK-less broadcast.
Furthermore, we proposed a sim-to-real learning-based method,
wherein a rate control policy is obtained
using DRL in the learning phase, and subsequently,
the policy is applied to the real environment as the application phase.
Numerical results confirmed that using the proposed sim-to-real DRL framework
can alleviate reception failures that could not be avoided
by the rule-based method.

\section*{Acknowledgement}

This research and development work was supported by the MIC/SCOPE \#JP196000002.

\bibliographystyle{IEEEtran}
\bibliography{main}

\end{document}